# Electroweak results from HERA

## Zhiqing Zhang[1]

LAL, Univ Paris-Sud, CNRS/IN2P3, Orsay, France

**Abstract**. The inclusive neutral and charged current deep inelastic scattering processes at high $Q^2$ provide the main basis for electroweak studies at HERA. These studies include the determination of the *W* boson propagator mass and weak neutral current couplings of light quarks to the $Z^0$ boson. The polarised electron or positron beam at HERA II gives rise to additional sensitivities for these studies. Exceptions also exist; this is the case for the single *W* production.

## 1. Introduction

The HERA electron-proton collider has come to an end after more than fifteen years of successful data taking. The first phase (HERA I) started in 1992 and lasted for eight years. In HERA I, the two general purpose detector experiments H1 and ZEUS have each collected an integrated luminosity of about 100pb$^{-1}$ in the collision mode of positron-proton ($e^+p$) and about 20pb$^{-1}$ in electron-proton ($e^-p$). The second phase (HERA II) covers the period of 2003-2007. In this period, the electron or positron beam at HERA II was polarised longitudinally for both H1 and ZEUS. In addition, the peak luminosity has been increased by a factor 3 due to a finer beam focusing around the interaction points and to larger beam currents. The total integrated luminosity per experiment taken in several data periods in both positive and negative polarisation modes is around 200pb$^{-1}$ for $e^+p$ as well as for $e^-p$.

At HERA, both neutral and charged current (NC and CC) processes are produced. The NC and CC cross sections are measured in a large kinematical domain covering more than 4 orders of magnitude both in $Q^2$, the negative four-momentum transfer squared, and in Bjorken *x*. The inclusive NC and CC cross sections are not only a primary source of data from which the parton distribution functions (PDFs) are extracted, but also provide basis for electroweak (EW) studies.

In the next section, a few selected EW results from HERA are presented. Most of them are based on the published HERA I data and part of the analysed HERA II data.

## 2. Electroweak results from HERA

### 2.1. *Unified electroweak forces and W propagator mass*

The NC process is dominated at low $Q^2$ by electromagnetic interactions with γ exchange in *t* channel. At high $Q^2$ additional contributions arise from $Z^0$ exchange and $\gamma Z^0$ interference. The CC process proceeds via weak interactions with a *W* boson only in *t* channel. The preliminary measurement of single differential NC and CC cross sections dσ/d$Q^2$ [1][2][3][4] based on part of HERA II data is shown in Fig. 1. Interestingly, the

---
[1] On behalf of the H1 and ZEUS Collaborations.

HERA data show that, at $Q^2$ values $\sim M_W^2$ and $M_Z^2$, the NC and CC cross sections become comparable, demonstrating the unified electromagnetic and weak coupling strengths in deep inelastic scattering processes. The residual difference between NC and CC and between $e^+p$ and $e^-p$ is understandable because of the up and down quark flavour asymmetry and the different helicity factors involved.

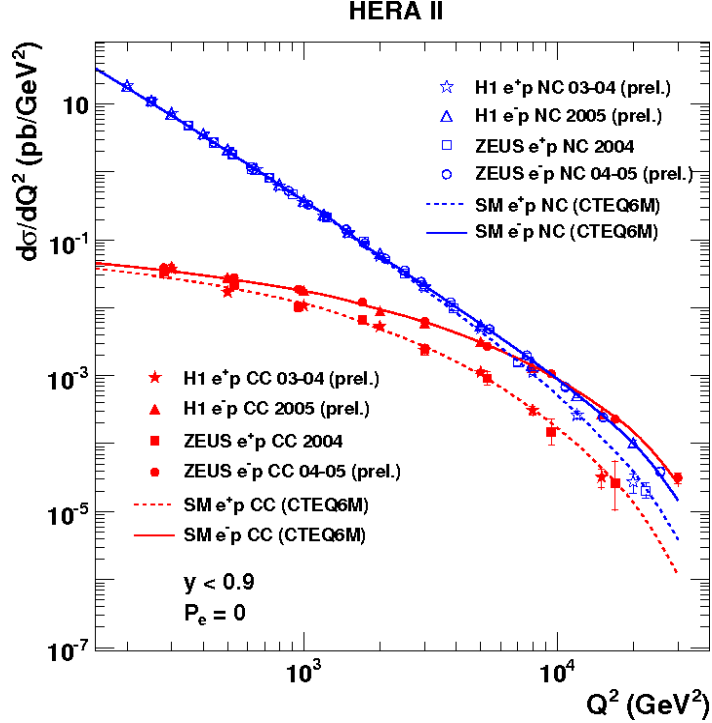

**Figure 1.** The $Q^2$ dependences of the NC and CC cross sections $d\sigma/dQ^2$ shown for the preliminary 03-05 $e^+p$ and $e^-p$ data. The inner and outer error bars represent the statistical and total errors respectively. The results are compared with the corresponding SM expectations determined from the CTEQ6M parameterisation [6]

The CC cross sections are measured to be much flatter than the NC cross section due to the large $W$ boson mass. The CC cross sections at HERA can thus be used to determine the propagator mass of the $W$ boson exchanged in the space-like regime, providing an independent measurement complimentary to other direct measurements performed using the real $W$ boson produced singly at the Tevatron and in pairs at LEP-2. The typical precision obtained with the HERA I data, in a model independent way, is around 1.8GeV [5]. Within the Standard Model (SM) due to constraints between different SM parameters, the precision is increased to about 200MeV [5]. The analysis of the full HERA II data should significantly improve the precision. This is because the main sensitivity arises from the larger CC cross section measured with the $e^-p$ data set, which has had a 10-fold increase at HERA II.

*2.2. High $Q^2$ NC cross sections and parity violation*

The NC cross section at low $Q^2$ is dominated by pure photon exchange. One way of accessing EW effects at high $Q^2$ is to measure the polarisation cross section difference. Indeed, the polarisation asymmetry measures a product of vector and axial-vector couplings and at HERA is sensitive to parity violation at spatial dimensions down to $10^{-18}$m. The first measurement of a NC polarisation asymmetry [7], performed at $Q^2 \sim$ 1GeV$^2$, was crucial in establishing the SM theory.

Using the preliminary NC cross sections measured with the first HERA II data set taken in 2003-2005 by both H1 and ZEUS, a combined polarisation asymmetry is obtained [8] and shown in Fig. 2. The asymmetry is well described by the SM predictions as obtained from the H1 [9] and ZEUS [10] QCD fits.

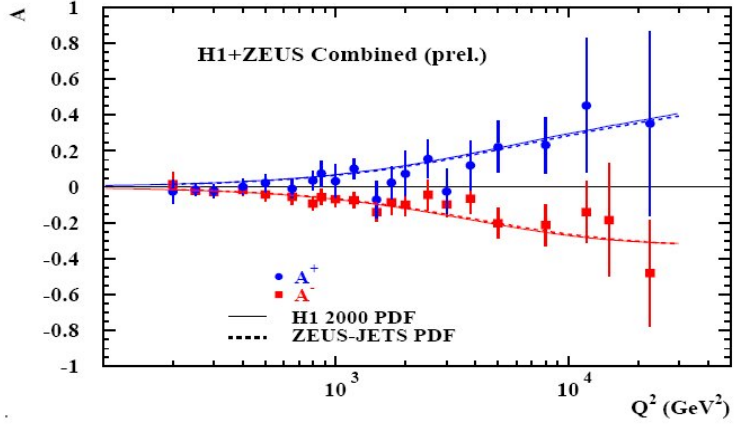

**Figure 2.** Measurements of the polarisation asymmetries $A^{\pm}$ based on a preliminary combination of the H1 and ZEUS data. The error bars denote the total uncertainty which is dominated by the uncorrelated error contributions. The curves describe the theoretical predictions in NLO QCD as obtained in fits to the H1 inclusive data and to the inclusive and jet ZEUS data, respectively. Both fits have been performed using the unpolarised HERA I data.

*2.3. Total CC cross sections and right-handed current*

The total CC cross sections have been measured by both H1 and ZEUS using their unpolarised HERA I data and polarised data taken in 2003-2005 in slightly different phase space. The H1 [2][9][11] and ZEUS [3][4][12] measurements, in a common phase space $Q^2>400\text{GeV}^2$ and $y<0.9$, are compared in Fig. 3 with SM expectations from CTEQ6D [6] and MRST [14]. The linear dependence on $P_e$ of the CC cross sections within the SM is expected as the $W$ boson interacts only with $e_L^-$ and $e_R^+$. A straight line fit to these cross sections is sensitive to exotic right-handed current additions to the SM lagrangian. Assuming the equal coupling strength between the right and left-handed currents and a light right-handed neutrino mass, the lower mass limits set on $W_R$ at 95% confidence level (CL) are 208 and 186GeV based on H1 $e^+p$ [13] and $e^-p$ [2] data respectively.

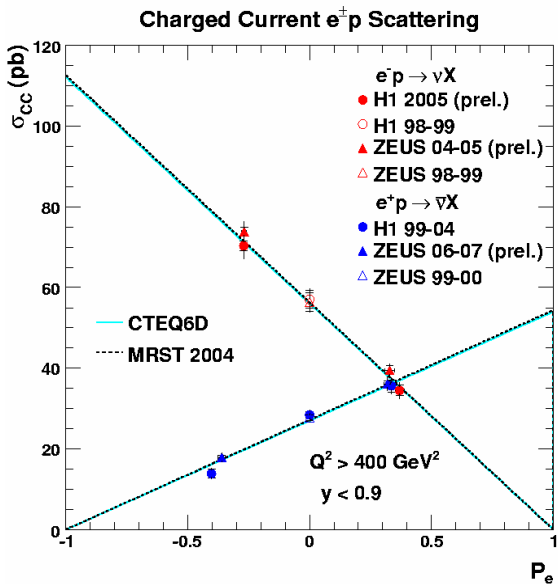
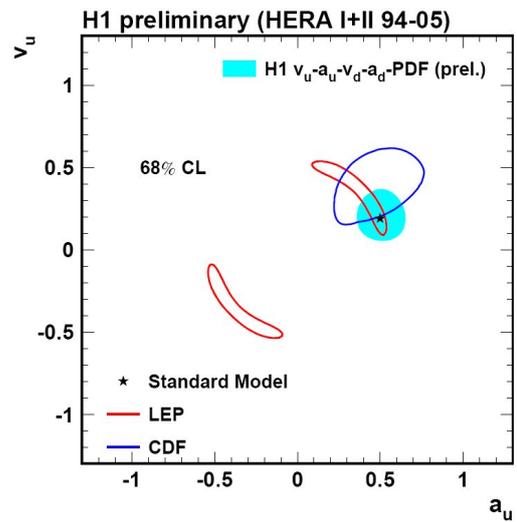

**Figure 3.** The dependence of the $e^{\pm}p$ CC cross sections on the lepton beam polarisation $P_e$. The inner and outer error bars represent the statistical and total errors respectively. The data are compared to SM predictions based on the CTEQ6D and MRST 2004 parameterisations.

**Figure 4.** Preliminary result at 68% CL on the weak neutral current couplings of $u$ quark to the $Z^0$ boson determined in a combined EW-QCD fit. The HERA result can be compared with those determined by the CDF experiment and from $e^+e^-$ measurements at the $Z^0$ resonance by LEP/SLC.

## 2.4. New sensitivity to light quark couplings to $Z^0$

At HERA, the first determination of the vector and axial-vector couplings of the light quarks $u$ and $d$ to the $Z^0$ boson has been obtained previously by H1 [5] in a combined EW-QCD fit using its published precision HERA I data at low $Q^2$ and NC and CC cross sections at high $Q^2$. The polarised electron or positron beam at HERA II has introduced additional sensitivity to these couplings, in particular to the vector ones. Indeed, including the preliminary NC and CC cross sections based on a subsample of the HERA II data taken in 2003-2005, new combined EW-QCD fits performed by both ZEUS [15] and H1 [16] have improved the determination of the couplings with better precision. The $u$ quark couplings determined by H1 are shown in Fig. 4 in which the determination is compared with those from the Tevatron and LEP/SLC.

## 2.5. Production of single W and a first determination of W polarisation fractions at HERA

Based on the full HERA I+II high energy data sample of 482 pb$^{-1}$ collected by H1 in the period 1994-2007, the single $W$ production cross section is measured [17] using events with an isolated electron or muon in the final state (thus from decay modes $W \rightarrow e/\mu/\tau + \nu$) and large missing transverse momentum. The result, $\sigma_W$ =1.23±0.25$_{stat}$±0.22$_{sys}$, is a measurement of 5$\sigma$ significance and is in good agreement with the SM expectation of 1.31±0.20$_{th.sys}$. The measurement of $W$ boson polarisation fractions makes use of the cos$\theta^*$ distributions in the decay $W \rightarrow e/\mu+\nu$. The angle $\theta^*$ is defined as the angle between the $W$ boson momentum in the lab frame and that of the charged decay lepton in the $W$ boson rest frame. The cos$\theta^*$ distribution has three components proportional to the left-handed fraction $F_-$, the longitudinal fraction $F_0$ and the right-handed fraction $F_+$ respectively. Only two of them are independent as $F_-+F_0+F_+=1$ and can be extracted in a fit to the measured distribution. The result of the fit is shown in Fig. 5.

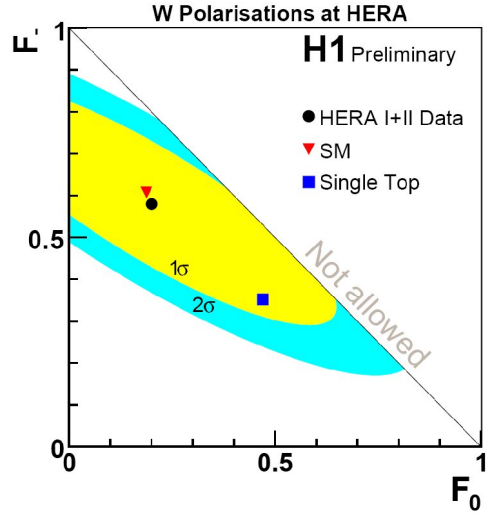

**Figure 5.** The fit result for simultaneously extracted left-handed $F_-$ and longitudinal $F_0$ $W$ boson polarisation fractions (point) at 1 and 2$\sigma$ CL (contours). Also shown are the values from the SM predictions (triangle) and anomalous single top production via flavour changing neutral current.

**Summary**

A few recent electroweak results from HERA have been reported. The results are mostly obtained using part of the full HERA data. The polarisation and the large luminosity data sample from HERA II will improve significantly most of the results presented here.